# Data is the Fuel of Organizations: Opportunities and Challenges in Afghanistan

Abdul Rahman Sherzad

Lecturer at Computer Science Faculty of Herat University, Afghanistan

Ph.D. Student at Technische Universität Berlin, Germany

absherzad@gmail.com || absherzad@mailbox.tu-berlin.de

## Abstract

In this paper, the author at first briefly outlines the value of data in organizations and the opportunities and challenges in Afghanistan. Then the author takes the Kankor (National University Entrance Exam) data, particularly names of participants, locations, high schools and higher education institutions into account and explains how these data, that organizations in Afghanistan do not use for anything, can be useful in several cases and areas. The application of these data is shown through cases such as Auto filling missing values, identifying names of people, locations, and institutions from unstructured text, generating fake data to benchmark the database and web application performance and appearance, comparing and matching high school data with Kankor data, producing the top-n male and female names very common in Afghanistan or province-wise, and the data mining application in education and higher education institutions.

## 1  Kankor in Afghanistan

To enter higher education, high school graduates are required to pass the National University Entrance Exam (Kankor, from the French word Concours; English: "Contest") . Kankor is held every year, mainly in the capital and large provinces.

The Kankor exam comprises 160 questions about subjects taught at high school mainly in grades 10 to 12 categorized into the following categories: Mathematics, Natural Science, Social Science, and Languages.

Kankor questions are provided both in Pashtu and Dari, in Pashtu and Dari speaking regions respectively. All the questions are multiple choice, and generally, the participants have two to three hours to answer and choose the proper options. Usually, correct answers are worth one to three points. The maximum number of points are between 320 and 370, but more important is the minimum number of points needed to qualify for a university slot. Applicants can choose five favorite fields of study, and each field of study requires a certain score.  In general, admission into popular fields of study like medicine, engineering, computer science, economics, and political sciences and into institutes in the capital and main cities requires higher scores. However, if a student does not score high enough for any of his/her chosen fields of study, he/she is dropped altogether and is not assigned any field of study, a result called result-less (benatedja).

The Kankor candidates could opt for ten fields of study they favor, and for one location for each field of study. They also had three chances to participate in Kankor in lifetime until 2011. Since then, they can opt for only five fields of study and have only two chances in the lifetime.

This paper (Sherzad, 2017) explains in detail the education and higher education system in Afghanistan as well as discusses the challenges and suggestions for improvement, and this paper (Sherzad, 2016b) provides comprehensive information and analyses of Kankor in Afghanistan and Kankor Data.

## 2   Data is Power

A car without fuel cannot be driven; a mobile, a laptop or a PC without power cannot be used; a website without feeding won't have any visitors; likewise, an organization without data will not stand and cannot be survived.

A decade or two ago fuels/oil were considered the most significant resource in the world. Then, water attracted attention of the world. Nowadays, data is considered the fuel of any country, organization and finally the future – as Dave Coplin, Microsoft Chief Envisioning Officer at Xerocon 2016 said (Heber, 2016), "Data is the fuel of our future. When you have lots of data, it changes things".

Data is power and without data it is not feasible to produce facts and figures, insights, analyses, mining and finally decisions cannot be made – the term "data" exists and plays the crucial role in **Data** Analyses, Knowledge Discovery from **Data** (KDD), **Data** Mining, **Data** Science, and Big **Data**. It is the data that enables organizations to explain the past and predict the future through data science and business intelligence tools.

## 3  Kankor Data and Applications

Since 2003, around 1.7 million eligible high school graduates have attended Kankor. Their First Names, Family Names (Last Names), Father Names, and Grand Father Names are recorded in the Kankor system. However, these data are not used at all by the educational institutions and other organizations.

Prior to performing any data analysis activities, such as exploring and visualizing the data, building a data mining model, or even generating reports, it is often required to transform and cleanse the data.

Study of Kankor dataset by the author this paper reveals that values contain leading or trailing spaces and even blank spaces in the middle that are unnecessary, or they have multiple embedded space characters (Unicode character set values 32 and 160), or nonprinting characters (Unicode character set values 0 to 31, 127, 129, 141, 143, 144, and 157). The unnecessary spaces and characters can sometimes cause unexpected results when data is sorted, filtered, matched or searched.

Moreover, some values are indistinguishable such as, عبدالوحید and عبدالوحيد. But in the former عبدالوحید for the letter Yeh (ی) Arabic Letter Farsi Yeh with Unicode 06CC is used. However, in the latter عبدالوحيد for the letter Yeh (ي) Arabic Letter Yeh with a completely different Unicode 064A is used.

To remove these nonessential spaces and characters, a combination of text transformations such as, Trim, Clean, Replace, and Substitute is used.

Taking these basic data into account leading us to the following practical examples and applications:

## 3.1 Top-n Male and Female Names in Afghanistan

The 1.7 million Kankor participants data include of 65,000 unique male and female First Names; more than 29,000 unique Family Names; more than 61,000 unique Father's Names, and finally more than 65,000 unique Grand Father's Names.

The Kankor participants First Names represent almost all the common names used in Afghanistan. Likewise, the Family Names are representing all the family names. Moreover, Father's Names and Grand Father's Names represent almost all the male names both stylish and classic.

All these names can be used to find the top-n male and female names largely used by the citizens in Afghanistan. After performing basic transformation on the data, primary analysis shows more than 11,000 Kankor male participants were named ذبیح الله (Zabiullah) and more than 9,000 female participants were named فاطمه (Fatemeh) across the country, as reflected (see **Figure 1**). Furthermore, the Province Names dimension/attribute can be added as part of the analysis to find the top-n male and female names province-wise.

| First Name (Male) | ASCII Format | Participants | First Name (Female) | ASCII Format | Participants |
|---|---|---|---|---|---|
| ذبیح الله | Zabiullah | 11,388 | فاطمه | Fatemeh | 9,346 |
| نجیب الله | Najibullah | 9,934 | فرشته | Freshteh | 6,993 |
| عبدالله | Abdullah | 9,558 | مریم | Maryam | 6,847 |
| سمیع الله | Samiullah | 8,261 | زهرا | Zahra | 6,304 |
| حمیدالله | Hamidullah | 8,129 | فرزانه | Farzana | 5,013 |
| روح الله | Rohullah | 7,387 | مرسل | Morsal | 4,395 |
| احسان الله | Ehsanullah | 7,167 | تمنا | Tamana | 4,251 |
| نقیب الله | Naqibullah | 6,823 | مرضیه | Marzia | 4,189 |
| رحمت الله | Rahmatullah | 6,737 | سمیرا | Samira | 3,842 |
| اسدالله | Asadullah | 6,461 | خاطره | Khatereh | 3,727 |
| حبیب الله | Habibullah | 6,200 | شکریه | Shokria | 3,609 |
| نعمت الله | Nematullah | 5,990 | معصومه | Masoma | 3,378 |
| عصمت الله | Esmatullah | 5,912 | نرگس | Nergis | 3,320 |
| حسیب الله | Hasibullah | 5,889 | شبانه | Shabana | 3,268 |
| وحیدالله | Wahidullah | 5,832 | فریده | Farideh | 3,210 |
| صفی الله | Safiullah | 5,603 | شگوفه | Shagofeh | 3,185 |
| مصطفی | Mustafa | 5,330 | آرزو | Arezo | 3,143 |
| محمد | Mohammad | 5,057 | مژگان | Mozhgan | 3,139 |
| عبیدالله | Obaidullah | 4,910 | نیلوفر | Nilofar | 3,033 |

**FIGURE 1: TOP 20 MALE AND FEMALE NAMES IN AFGHANISTAN.**

It is worth mentioning, that even these numbers illustrated in **Figure 1** are not very accurate because in Dari/Persian same names can be written in different variations such as the name ذبیح الله (Zabiullah) can be written as "ذ بیح الله" with a space after the first character, "ذبیحالله" without any space and with the two words connected or other variations that are not identical while performing analysis.

In Dari/Persian a character may take different shapes depending to its position in the string and the character which comes before or after it. For instance, when the character Aleph (or alef or alif; Persian/Dari: الف) follows the character Re (ر) such as in "زهرا" it is written separately and not connected with its previous character. But when the character الف follows the character Fe (ف) such as in "فاطمه" it is connected with its previous character.

| Name | Len | Variations | Len | Matched? | Comment |
|---|---|---|---|---|---|
| فاطمه | 5 | فاطمه | 6 | FALSE | One unnecessary trailing space |
| فاطمه | 5 | فا طمه | 6 | FALSE | One unnecessary space in middle |
| فاطمه | 5 | فا طمه | 7 | FALSE | Two unnecessary spaces in middle |
| فاطمه | 5 | فاطمه | 7 | FALSE | Two unnecessary leading spaces |
| ذبیح الله | 9 | ذبیح الله | 11 | FALSE | One unnecessary leading space |
| ذبیح الله | 9 | ذ بیح الله | 10 | FALSE | One unnecessary space after 1st char |
| ذبیح الله | 9 | ذبیح الله | 10 | FALSE | Two unnecessary spaces in middle |
| ذبیح الله | 9 | ذبیحالله | 8 | FALSE | Two words are connected |

Transformed

| Name | Len | Variations | Len | Matched? | Comment |
|---|---|---|---|---|---|
| فاطمه | 5 | فاطمه | 5 | TRUE | Trailing space was trimmed |
| فاطمه | 5 | فا طمه | 6 | FALSE | Space in middle could not be trimmed |
| فاطمه | 5 | فا طمه | 6 | FALSE | Only one of middle spaces were trimmed |
| فاطمه | 5 | فاطمه | 5 | TRUE | Leading spaces were trimmed |
| ذبیح الله | 9 | ذبیح الله | 9 | TRUE | Leading spaces were trimmed |
| ذبیح الله | 9 | ذ بیح الله | 10 | FALSE | Space after 1st char could not be trimmed |
| ذبیح الله | 9 | ذبیح الله | 9 | TRUE | Only one of middle spaces were trimmed |
| ذبیح الله | 9 | ذبیحالله | 8 | FALSE | Nothing happened |

**FIGURE 2: THE BUILT-IN TRIM FUNCTION CANNOT REMOVE THE SINGLE UNNECESSARY SPACE FROM DARI/PERSIAN TEXT.**

When a character is unconnected to its previous character, that does not mean that there should be a space between the two characters. In fact, spaces are not allowed between unconnected characters within one word. An example is فاطمه (there is no space between its unconnected characters). If there is only one space between the unconnected characters of a word (for example in فا طمه in which there is one additional space after the character الف), the built-in trim function cannot remove that space, because the trim function was designed to remove two or more connected spaces. Thus, the names فاطمه and فا طمه will not be identical, even after the trim transformation is carried out because the built-in trim function cannot remove the single unnecessary space in the middle of strings, as illustrated in the above (see **Figure 2**) for further clarification.

After addressing the above challenge systematically, analysis shows changes in the number under the "Participants" attribute. Prior to addressing the challenge, the name ذبیح الله (Zabiullah) appeared 11,388 times, but after addressing the above challenge it appeared 11,450 times. Also, the name حمیدالله (Hamidullah) ranked higher than the name سمیع الله (Samiullah) as shown in the following figure (see **Figure 3**):

| First Name (Male) | ASCII Format | Participants | First Name (Female) | ASCII Format | Participants |
|---|---|---|---|---|---|
| ذبیح الله | Zabiullah | 11,450 | فاطمه | Fatemeh | 9,372 |
| نجیب الله | Najibullah | 9,966 | فرشته | Freshteh | 7,003 |
| عبدالله | Abdullah | 9,658 | مریم | Maryam | 6,856 |
| حمیدالله | Hamidullah | 8,581 | زهرا | Zahra | 6,307 |
| سمیع الله | Samiullah | 8,294 | فرزانه | Farzana | 5,020 |
| روح الله | Rohullah | 7,423 | مرسل | Morsal | 4,397 |
| احسان الله | Ehsanullah | 7,203 | تمنا | Tamana | 4,251 |
| نقیب الله | Naqibullah | 6,837 | مرضیه | Marzia | 4,194 |
| اسدالله | Asadullah | 6,769 | سمیرا | Samira | 3,843 |
| رحمت الله | Rahmatullah | 6,755 | خاطره | Khatereh | 3,736 |
| وحیدالله | Wahidullah | 6,269 | شکریه | Shokria | 3,609 |
| حبیب الله | Habibullah | 6,209 | معصومه | Masoma | 3,386 |
| نعمت الله | Nematullah | 6,028 | نرگس | Nergis | 3,329 |
| عصمت الله | Esmatullah | 5,936 | شبانه | Shabana | 3,271 |
| حسیب الله | Hasibullah | 5,909 | فریده | Farideh | 3,218 |
| صفی الله | Safiullah | 5,615 | شگوفه | Shagofeh | 3,198 |
| مصطفی | Mustafa | 5,330 | مژگان | Mozhgan | 3,153 |
| عبیدالله | Obaidullah | 5,120 | آرزو | Arezo | 3,146 |
| محمد | Mohammad | 5,057 | نیلوفر | Nilofar | 3,061 |
| نصیراحمد | Nasir Ahmad | 4,972 | راضیه | Razia | 3,028 |

**FIGURE 3: TOP 20 MALE AND FEMALE NAMES AFTER PROPER PREPROCESSING AND TRANSFORMATION PROCESSES WERE PERFORMED.**

Likewise, the very common Family Names across the country in Afghanistan or province-wise can be found.

## 3.2 Autofill Missing Gender Values Using Identical Names

Out of 1.7 million records of Kankor data, for around 200,000 records the gender label is missing. This situation poses a challenge for data quality and completeness and should be addressed systematically through a structured process. The current solution is based on the fact that there are many Kankor applicants with the same first names e.g. more than 11,000 were named ذبیح الله (Zabiuallah), as described in the previous section. It is assumed that the gender for at least one of the identical first names is identified in the master list. That gender value is used as a reference to autofill the gender for the identical names whose gender value is missing, as it is reflected in the following figure.

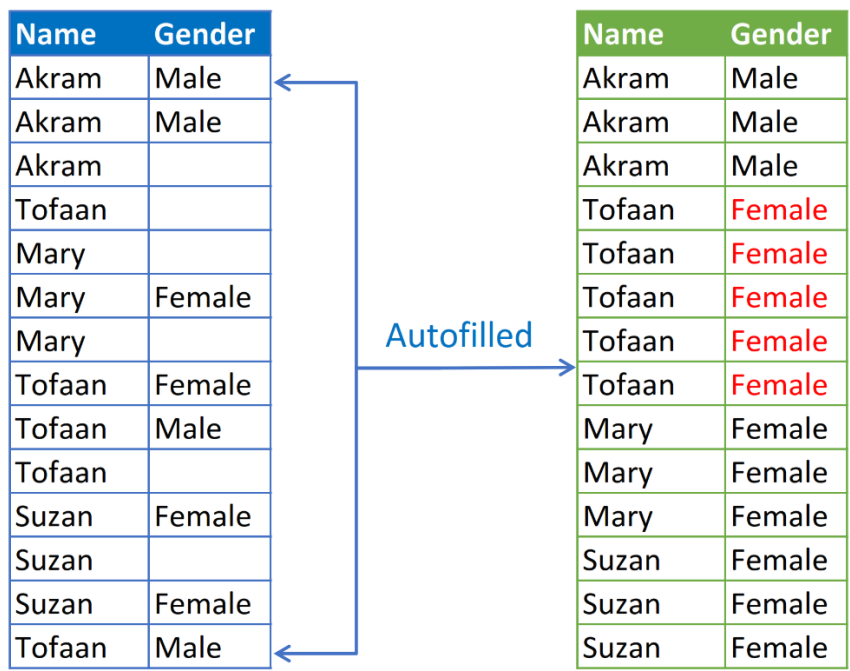

**FIGURE 4: AUTOFILL MISSING GENDER VALUES FROM EXISTING DATA USING IDENTICAL NAMES.**

It is worth mentioning that in some rare cases there are first names used both for men and women. In the Kankor data consider the following scenario: A female Kankor candidate, named طوفان (Tofaan), has a female label in the gender column. There are records of male candidates with the same names with male labels or the labels for their gender column are missing. Because the names are identical, the approach described above may take the gender value of the female candidate and autofill for the missing values for the male candidates, which is false (see **Figure 4**). This situation raises a challenge for the labeling of the missing values for the gender column.

To address this predicament, other approaches were used to guarantee that the algorithm will autofill the missing values properly. This approach divides the dataset into two datasets, one dataset contains only instances with missing gender value, and the other dataset contains identical names of those with the maximum number of occurrences of gender values. Then these two datasets are merged for comparison and based on the maximum number of occurrences of existing values in the second dataset the missing gender values in the first dataset are auto filled, and it also ensures the values already entered by the administrators are not overwritten, as shown (see **Figure 5)**.

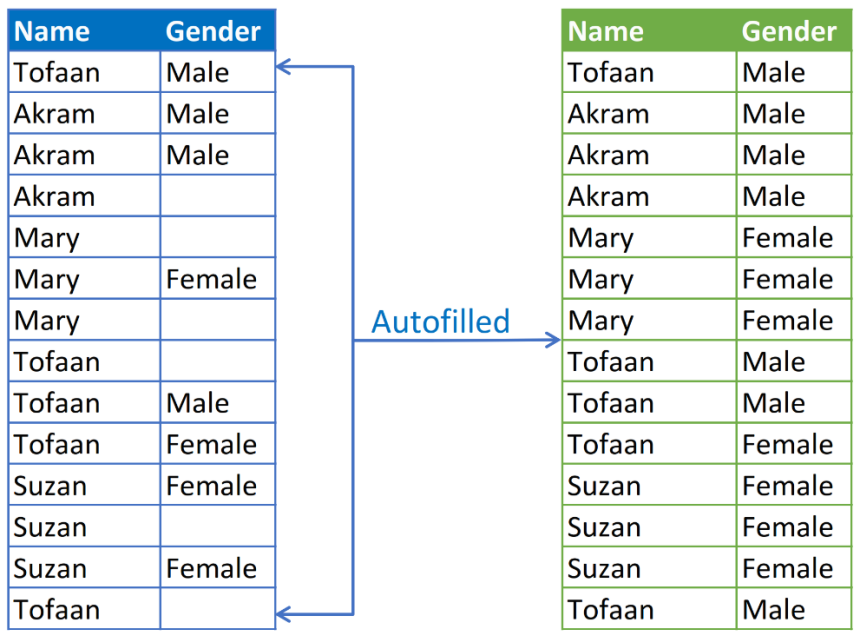

FIGURE 5: AUTO FILL MISSING GENDER VALUES BASED ON THE MAXIMUM NUMBER OF OCCURRENCES OF EXISTING VALUES.

It is worth mentioning that for some cases the High School Names indicate whether the school is for girls or boys and this attribute also can be considered to ensure auto filling the missing gender values are done more precisely.

## 3.3 Autofill Missing High School Geographical Location Values

The value for the location of high school column is missing for more than 600,000 records out of 1.7 million records of Kankor data. Likewise, the above technique used to autofill

P a g e 8 | 14

the missing gender values from existing data can be used to autofill the missing geographical location of high schools.

It is worth mentioning that in rare cases high schools share the same name across the country. Hence, there is a chance for two high schools to be located in different locations (say, one is located in Herat province and the other in Kabul province) but have the same name. In such a situation, the above-mentioned approach does not work and there is the possibility to misallocate values – In such a situation eyeballing by the administrators cannot allocate the missing values properly.

### 3.4 Audit and Match High School Data with Kankor Data

To find out the relation between high school marks and the success rate of candidates in the Kankor exam, it is required to merge and compare the candidates' high school marks with the Kankor data.

Since there are no common keys for these two datasets, the only solution is to use the combination of following common attributes for comparison: First Name, Father Name, Province, High School Name and Graduation Year.

Afghanistan consists of 34 provinces, hence, the values for the "Province" attribute are limited and can be rectified with basic preprocessing and transformation steps. The values for the "Graduation Year" are numeric and relatively straight-forward to be preprocessed and cleansed. However, the values for the candidates' First Names, Father Names and High School Names are tricky and cannot be rectified using the combination of clean, trim, replace and other preprocessing and transformation steps. The author of this paper proposes the following approaches addressing the issue of unnecessary spaces available in the First Names, Father Names, and High School Names:

1. **Approximate String Matching:** Approximate string matching, conversationally referred to as "fuzzy string searching", is a technique of finding strings that match a pattern approximately (rather than exactly). In this case, the approximate string matching is not a great solution because of the following reasons: there are names with equal number of similar characters that are completely different such as "رامز" and "مزار"; or there are names with one different character that are still completely different names such as جمال الدین (**J**amaluddin) and کمال الدین (**K**amaluddin). However, the approximate string matching often produces perfect or very high similarity score in such cases which is not correct, as it is reflected (see **Figure 6**). On the other hand, this technique is very time and resource consuming.

FIGURE 6: OFTEN APPROXIMATE STRING MATCHING PRODUCES HIGH SIMILARITY SCORE FOR COMPLETELY DIFFERENT NAMES.

2. **Exact String Matching:** This technique is the simplest version of string matching. However, in our case the exact string matching does not match all those same names written in different variations e.g. consisting additional unnecessary spaces such as فاطمه without any additional spaces and فا طمه with an additional space as described in the *Top-n male and female names in Afghanistan* section above.

3. **Tailored Trim Method:** All the existing trim methods are designed to remove all the leading and trailing spaces from a given string. Additionally, depending on the programming languages they can also remove only additional spaces (more than one consecutive spaces) from the middle of a given string. But in the case Dari/Persian there are many situations and cases (one) additional spaces are available in different positions of the names that causes the names do not match and the trim methods cannot detect to trim those spaces from the names. Hence, a customized trim method is required to be designed to correctly identify and trim those one spaces as additional spaces from the names to be ready for matching. Designing such a method is complex and requires contribution of expertise from linguistic and other interdisciplinary.

4. **Strip all Spaces:** In this method, initially all spaces from First Names, Father Names, and High School Names of both datasets were completely removed. Then through exact string matching all same instances from the two datasets were compared and matched precisely, as it is shown in the following figure (see **Figure 7**). This method also has been used to produce the top-n male and female names more precisely.

| Name | Len | Variations | Len | Matched? | Comment |
|---|---|---|---|---|---|
| فاطمه | 5 | فاطمه | 6 | FALSE | One unnecessary trailing space |
| فاطمه | 5 | فا طمه | 6 | FALSE | One unnecessary space in middle |
| فاطمه | 5 | فا طمه | 7 | FALSE | Two unnecessary spaces in middle |
| فاطمه | 5 | فاطمه | 7 | FALSE | Two unnecessary leading spaces |
| ذبيح الله | 9 | ذبيح الله | 11 | FALSE | One unnecessary leading space |
| ذبيح الله | 9 | ذ بيح الله | 10 | FALSE | One unnecessary space after 1st char |
| ذبيح الله | 9 | ذبيح الله | 10 | FALSE | Two unnecessary spaces in middle |
| ذبيح الله | 9 | ذبيحالله | 8 | FALSE | Two words are connected |

Spaces Stripped ↓

| Name | Len | Variations | Len | Matched? | Comment |
|---|---|---|---|---|---|
| فاطمه | 5 | فاطمه | 5 | TRUE | All spaces were stripped |
| فاطمه | 5 | فاطمه | 5 | TRUE | All spaces were stripped |
| فاطمه | 5 | فاطمه | 5 | TRUE | All spaces were stripped |
| فاطمه | 5 | فاطمه | 5 | TRUE | All spaces were stripped |
| ذبيحالله | 8 | ذبيحالله | 8 | TRUE | All spaces were stripped |
| ذبيحالله | 8 | ذبيحالله | 8 | TRUE | All spaces were stripped |
| ذبيحالله | 8 | ذبيحالله | 8 | TRUE | All spaces were stripped |

FIGURE 7: ALL SPACES WERE REMOVED FROM NAMES AND THEN EXACT STING MATCHING HAS BEEN USED FOR COMPARISON.

## 3.5 Generate Fake Data

Sometimes it is required to benchmark and measure the performance of databases or web applications with thousands and millions of proper manipulated fake data; or to test and evaluate the feel and look of web applications; or for other scenarios and circumstances. Around 1,000 male candidate names, 1,000 female candidate names, 1,000 family names, all the 34 province names, and more than 400 educational institution names were selected from Kankor data. Additionally, a script was written to convert them to their ASCII equivalent. Finally, these data were added as a "fa_AF" provider to "fzaninotto/Faker", a PHP library that generates fake data such as random male or female First Names and Family Names, Usernames, Emails, Websites, and other in many different languages.

### 3.6   Name Entity Recognition

Another great application of these data is in Text Mining, mainly, Name Entity Recognition (NER). Let's suppose you are asked to detect and identify "Person Names" within Persian/Dari unstructured text. The Kankor participants First Names, Family Names, Father Names, Grand Father Names can be considered as representative of all the names that are used in Afghanistan since they cover almost all the common. Most importantly, they are written and available into almost all possible different variations and formats including prefixes and suffixes. These names can be used as a rich dataset in the corpus for identifying the "Person Names" from large amounts of unstructured text.

Likewise, the Province, District, and Village Names can be used to identify the Locations from a given unstructured text.

Moreover, High School Names together with Educational Institutions Names can be used to identify almost all the Educational Institutions from a given unstructured text.

## 4   Educational Data Mining Applications

There are large amounts of data available for mining purposes in the context of Afghanistan, mainly, in education domain. But the methods that the educational institutions use to store and produce their data only enable them to achieve basic insights which are not useful for decision-makers or policy-makers and which do not help them to guess the future. Their main efforts are to generate (only) basic facts and figures (e.g., total number of students and teachers categorized by gender, location, and some other criteria). However, it turns out that these simple facts and figures do not help policy makers and educational institutions to improve the educational settings. For instance, they cannot be used to predict the right fields of study for high school graduates, or to identify first year university students who are at high risk of attrition or failure, or to recommend appropriate courses for university students.

It is very important to store detailed data and then to apply Data Mining techniques to explain and analyze the past such as what happened and why it happened as well as to predict the future such as what might happen and what should the organizations do.

From the results it can be concluded that there are potential opportunities for educational data mining application in the domain of Afghanistan's education systems (Sherzad, 2016a). Examples include:

- Predicting suitable fields of study for high school graduates prior to taking the exam.
- Alternatively, the data can provide the necessary tools to policy makers in shaping the education system by introducing specialized education at the high school level.
- Data also provides early warning systems to intervene in identifying first year university students who are at high risk of attrition caused by both capacity related limitations and other external obstacles.

Education Management Information System (EMIS) at Ministry of Education (MoE) and Higher Education Management Information System (HEMIS) at Ministry of Higher Education (MoHE) together could be appointed to provide the raw data for Data Mining applications to help discern patterns of abilities and behaviors which could be used to help educational institutions.

## 5 Conclusion

Lack of data availability and accessibility, lack of detailed data, and lack of experts are the main challenges preventing applicability of Data Mining and Educational Data Mining both at government and non-government organizations. More importantly, the concept of Data Mining is new, and research has not been done about its applications in Afghanistan. This study will be the beginning of a new era for other researchers and policy analysts both at the private and public sectors.